\documentclass[aps,prl,twocolumn,floatfix,showpacs]{revtex4}

\usepackage{graphics}
\usepackage{bm}

\def\PRA{{Phys.~Rev.~A} }

\def\JPB{{J.~Phys.~B} }
\def\PRL{{Phys.~Rev.~Lett.} }
\def\RMP{{Rev.~Mod.~Phys.} }
\def\JCP{{J.~Chem.~Phys.} }

\newcommand{\myscaleboxb}[1]{\scalebox{0.4}[0.4]{#1}}
\newcommand{\myscaleboxc}[1]{\scalebox{0.19}[0.19]{#1}}

\newcommand{\be}{\begin{equation}}
\newcommand{\bea}{\begin{eqnarray}}
\newcommand{\eea}{\end{eqnarray}}
\newcommand{\ee}{\end{equation}}

\begin{document}

\title{Polarization and ellipticity of high-order harmonics from
aligned molecules generated by linearly polarized intense laser pulses}

\author{Anh-Thu Le,$^1$ R.~R. Lucchese,$^2$, and C.~D. Lin$^1$}

\affiliation{$^1$Department of Physics, Cardwell Hall, Kansas
State University, Manhattan, KS 66506, USA\\
$^2$Department of Chemistry, Texas A\&M University, College Station,
Texas 77843-3255, USA}

\date{\today}

\begin{abstract}
We present theoretical calculations for polarization and ellipticity
of high-order harmonics from aligned N$_2$, CO$_2$, and O$_2$
molecules generated by linearly polarized lasers. Within the rescattering
model, the two polarization amplitudes of the harmonics are
determined by the photo-recombination amplitudes for photons emitted
parallel and perpendicular to the direction of the {\em same}
returning electron wave packet. Our results show clear
species-dependent polarization states, in excellent agreement with
experiments. We further note that the measured polarization ellipse
of the harmonic furnishes the needed parameters for a ``complete''
experiment in molecules.

\end{abstract}

 \maketitle

High-order harmonic generation (HHG) is  one of the most important
nonlinear processes that occur when atoms or molecules are placed in
an intense laser field \cite{krausz09}. Today these high harmonics
are used as convenient laboratory XUV or soft X-ray light sources,
as well as the sources of single attosecond pulses or attosecond
pulse trains \cite{Hentschel,Sasone,Remitter}. High harmonics are
emitted when laser-induced continuum electrons recombine with the
target ions. Since photo-recombination is a time-reversed process of
photoionization (PI), study of HHG from molecular targets offers
alternative means for probing molecular structure that have been
traditionally carried out using PI at synchrotron radiation
facilities. Gaseous molecules can be
given a periodic transient alignment
by a weak short laser pulse \cite{seideman}. By
studying HHG generated from such aligned molecules, information such
as molecular frame photoelectron angular distributions (MFPAD) for
PI from valence orbitals of molecules can be inferred. The goal of a
``complete experiment'' is to determine amplitudes and phases of all
dipole matrix elements. For linear molecules, this may be achieved
if measurements of MFPAD are carried out using elliptically
polarized lights \cite{Yagishita02,Lebech03}. For
photo-recombination, this means that one may obtain equivalent
information by examining the elliptical polarization of HHG from
aligned molecules.

Clearly if the gas is isotropically distributed, as for atomic or
unaligned molecular targets, due to the symmetry the emitted
harmonics are polarized parallel to the polarization of the driving
linearly polarized laser. For aligned molecules, a harmonic component
perpendicular to the laser polarization is expected to be present in
general \cite{lein-jmo05}. This requires that experiments be carried
out with a good level of molecular alignment in order to observe a
significant amount of the perpendicular harmonic component. It is
therefore not surprising that polarization measurements for emitted
harmonics were reported only very recently
\cite{levesque07,mairesse08,mairesse08b,nam08,jila09}. All these
experiments were carried out within the pump-probe scheme, where a
relatively weak, short laser pulse is used to impulsively align
molecules along its polarization direction, and after some delay
time, a second laser pulse is used to generate high-order harmonics.
We note that the commonly used strong-field approximation (SFA)
cannot be employed to interpret such experiments since it predicts
little or no ellipticity for emitted harmonics
\cite{levesque07,madsen09}.

In this Letter we report theoretical results for polarization and
ellipticity of HHG from aligned N$_2$, O$_2$, and CO$_2$ molecules.
Our results show very good agreement  with experimental measurements
\cite{jila09,levesque07,mairesse08} for harmonic orientation angles
and the reported large ellipticity for N$_2$ \cite{jila09}. The
calculations were carried out using the quantitative rescattering
theory (QRS) \cite{toru08,atle09,atle09-long,atle08,H2+} where the
complex induced dipole responsible for harmonic emission is
represented as a product of a returning electron wave packet and the
{\em laser-free} photo-recombination transition dipole,
\begin{equation}
D_{\parallel,\perp}(\omega,\vartheta)=W(E_k,\vartheta)d_{\parallel,\perp}(\omega,\vartheta).
\end{equation}
Here $\vartheta$ is the angle between the molecular axis and the
(probe) laser polarization direction, $E_k$ is the ``incident''
energy of the returning electron, and $\omega=I_p+E_k$ is the
emitted photon energy, with $I_p$ being the ionization potential.
The returning electron can recombine with the parent ion to emit a
photon with polarization in the parallel or perpendicular direction
to its motion, resulting in the two polarization components of the
emitted harmonics. Both of these complex transition dipoles
$d_{\parallel,\perp}$ are obtained from state-of-the-art molecular
photoionization code \cite{lucchese82,lucchese95} for each
fixed-in-space molecule. Note that we only need to consider the
harmonic components on the plane perpendicular to the propagation
direction of the driving laser, since only the harmonic emission
propagating along this direction can be efficiently phase matched.
As for the returning electron wave packet, we extract it from the
SFA \cite{atle09-long}. Eq.~(1) thus shows that the amplitude and
phase of the transition dipoles can be probed by studying HHG.

To compare with experiments, induced dipoles
$D_{\parallel,\perp}(\omega,\vartheta)$ from the fixed-in-space
molecules are coherently convoluted with the molecular alignment
distributions \cite{atle09-long,lein-jmo05,seideman07}. We note that
this alignment ``phase-matching'' tends to favor the parallel
component. In our simulations, the alignment distribution is
obtained from numerical solution of the time-dependent Schr\"odinger
equation within the linear rotor model for each molecular species
\cite{atle09-long,seideman}.  We use a 120 fs pump laser pulse with
an intensity of $3\times 10^{13}$ W/cm$^2$, and a 30 fs probe laser
pulse with an intensity of $2\times 10^{14}$ W/cm$^2$. Both pump and
probe lasers are of  800 nm wavelength. The rotational temperature
is assumed to be 100 K. These parameters were chosen to closely
match the experimental conditions of Zhou {\it et al} \cite{jila09}.
We vary the angle between pump and probe polarizations and use the
half-revival time delay for N$_2$ and O$_2$, and 3/4-revival for
CO$_2$.

\begin{figure}
\centering \mbox{\rotatebox{0}{\myscaleboxb{
\includegraphics{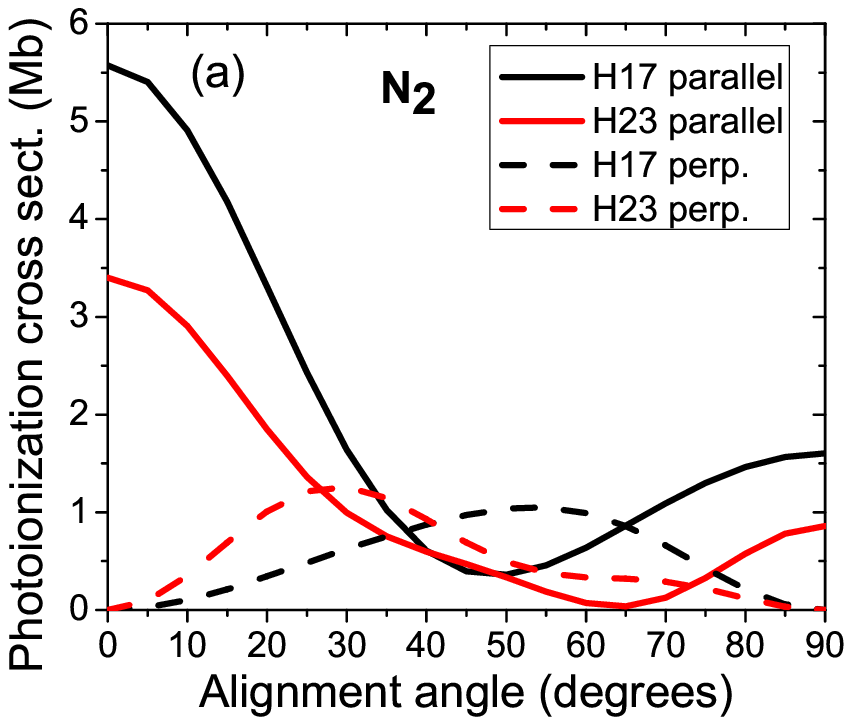}}}}
\centering \mbox{\rotatebox{0}{\myscaleboxb{
\includegraphics{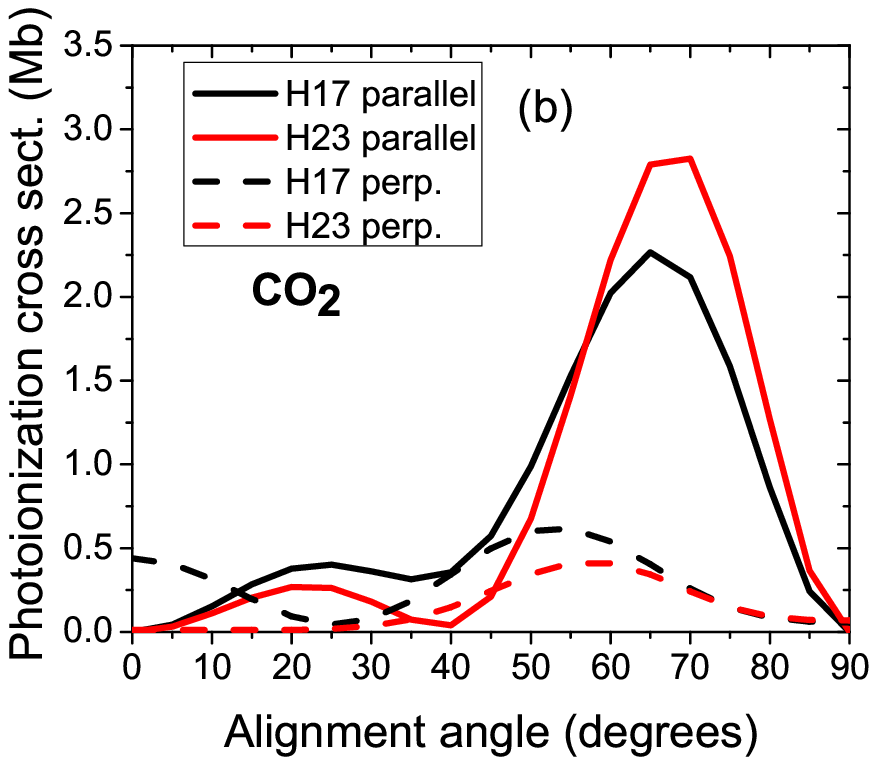}}}}
\centering \mbox{\rotatebox{0}{\myscaleboxb{
\includegraphics{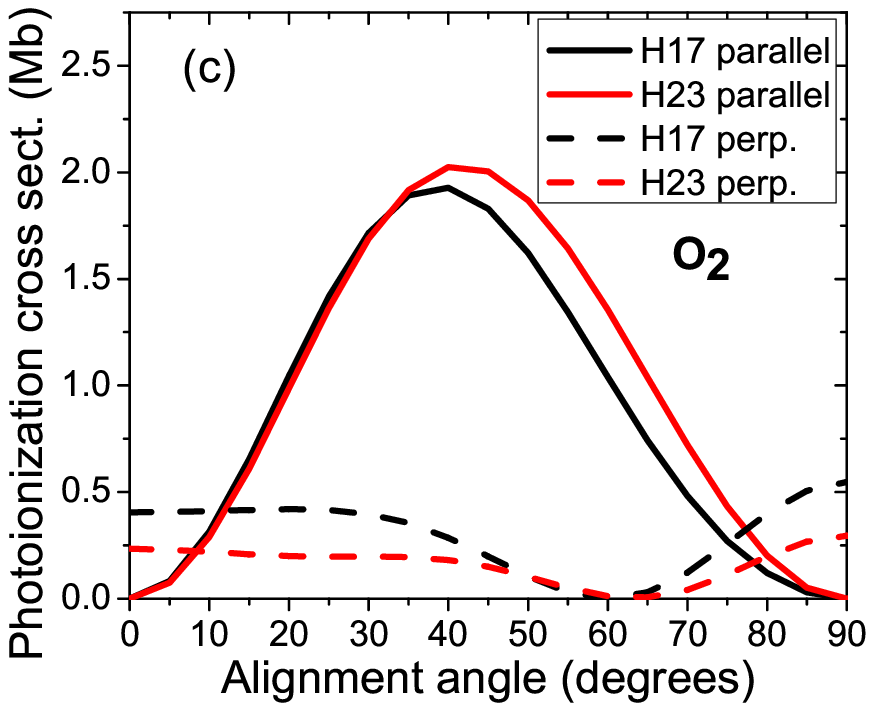}}}}
\caption{(Color online) Fixed-in-space molecular photoionization
differential cross sections, corresponding to the parallel and
perpendicular components of emitted harmonics H17 and H23 for N$_2$
(a), CO$_2$ (b), and O$_2$ (c).} \label{fig1}
\end{figure}

To understand the experimental measurements, we show in Fig.~1
fixed-in-space PI differential cross sections for the three species,
corresponding to the parallel and perpendicular components of
emitted harmonics H17 and H23. Note that these cross sections are
proportional to $|d_{\parallel}|^2$ and $|d_{\perp}|^2$,
respectively. In general, the perpendicular components are
smaller than the parallel components for all three targets.
Due to symmetry, the perpendicular harmonic component will vanish
after averaging over the alignment distribution if the angle between
pump and probe polarizations $\theta$ is $0^{\circ}$ or $90^{\circ}$
(see Fig.~2 below). From Fig.~1 we note that at intermediate angles
the cross section for the perpendicular component is quite comparable to
the parallel one for N$_2$ and CO$_2$. For O$_2$ the perpendicular
is always much smaller than the parallel one. Therefore one can
expect a small intensity ratio between perpendicular and parallel
components for O$_2$, but a larger ratio for both N$_2$ and CO$_2$.

\begin{figure}
\centering \mbox{\rotatebox{0}{\myscaleboxb{
\includegraphics{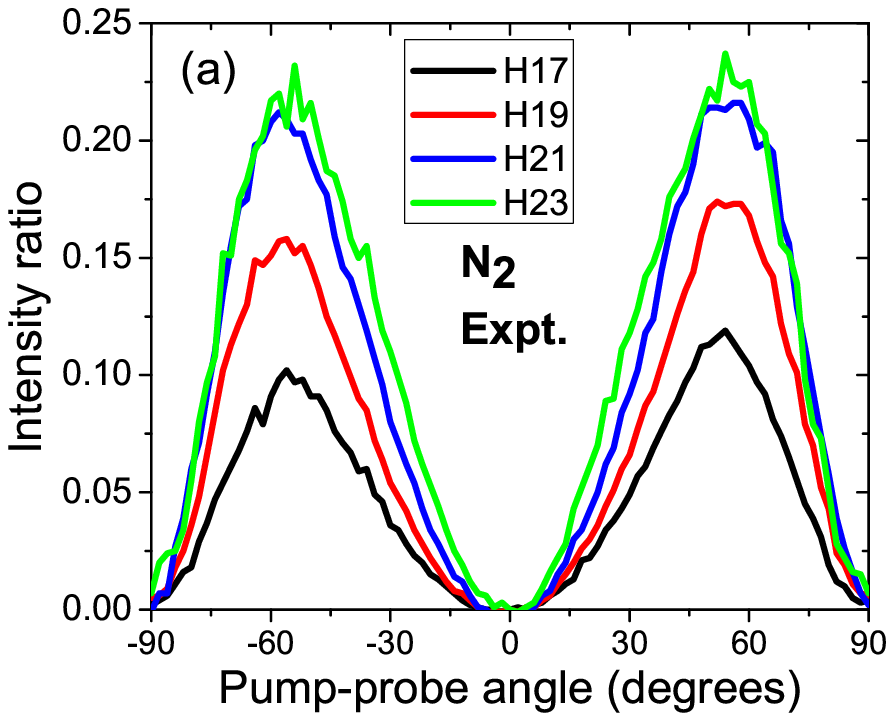}}}}
\centering \mbox{\rotatebox{0}{\myscaleboxb{
\includegraphics{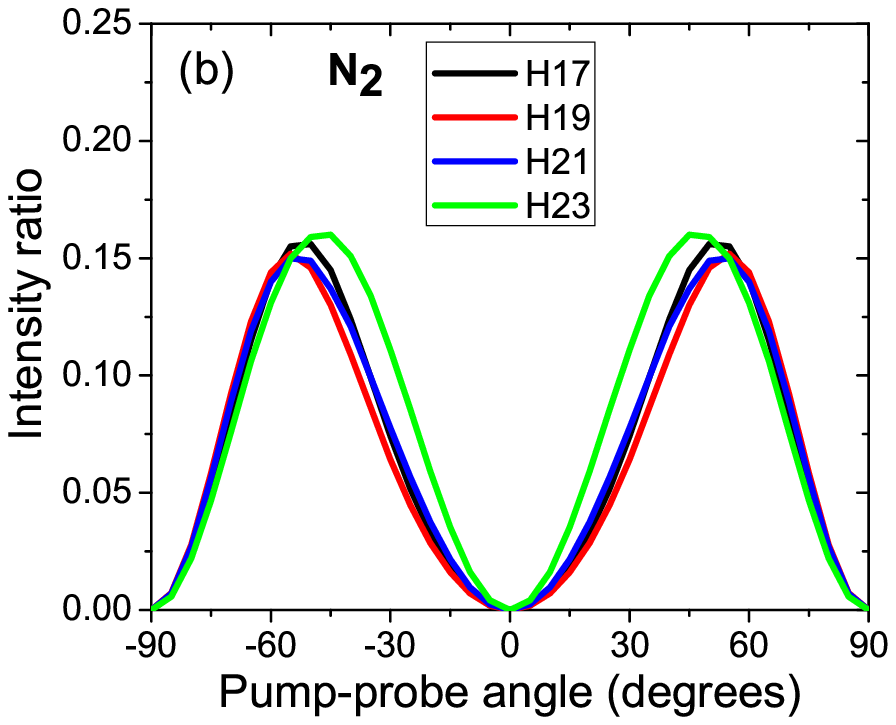}}}}
\centering \mbox{\rotatebox{0}{\myscaleboxb{
\includegraphics{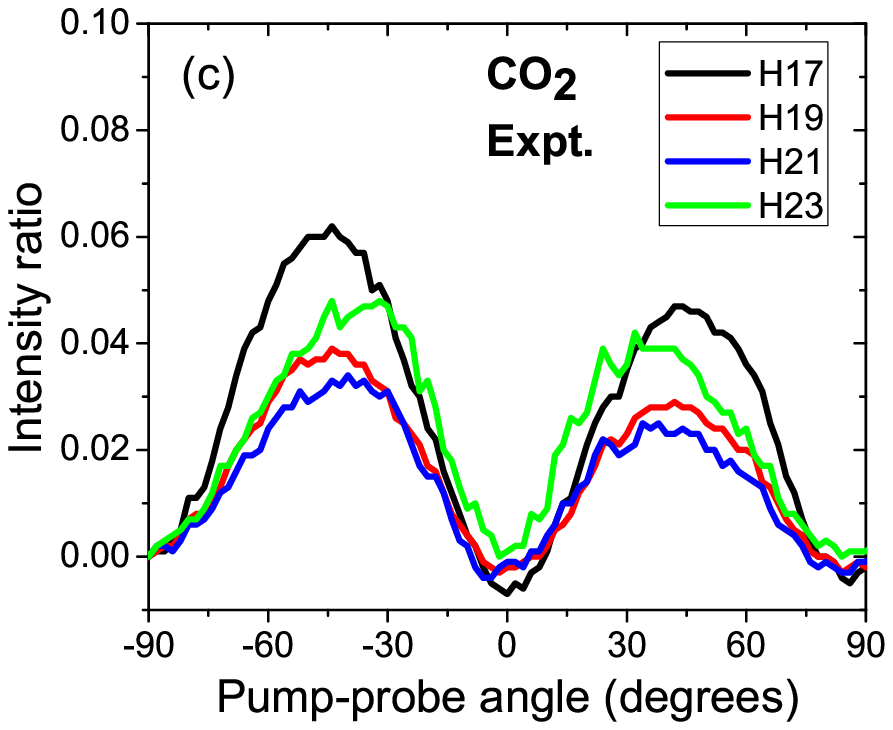}}}}
\centering \mbox{\rotatebox{0}{\myscaleboxb{
\includegraphics{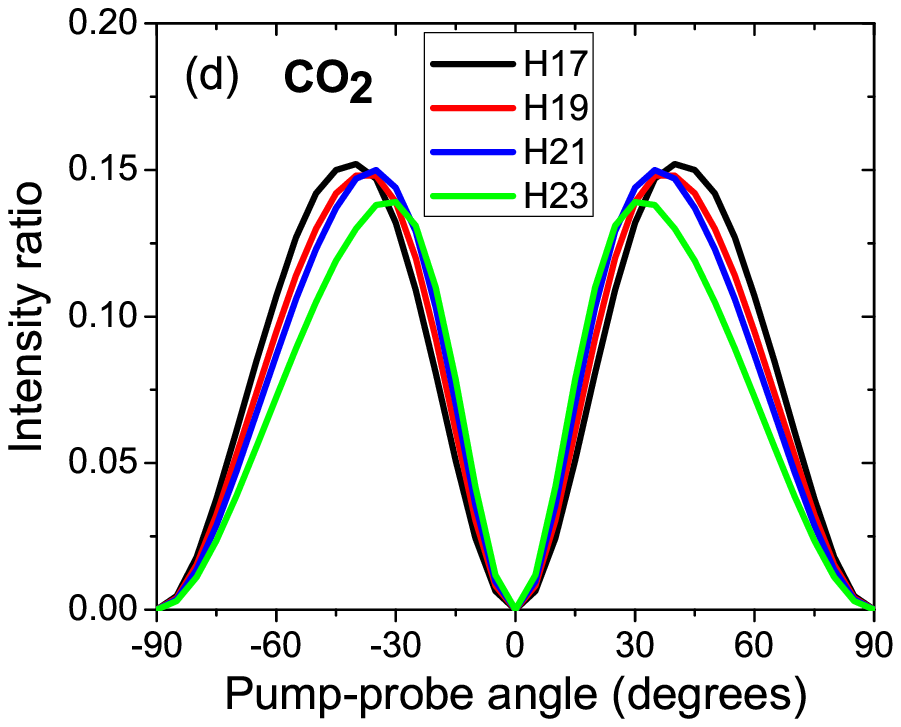}}}}
\caption{(Color online) Experimental (left panels) and theoretical
(right panels) intensity ratios between perpendicular and parallel
component of harmonic fields, as a function of alignment angle
between pump and probe polarization directions, for N$_2$ (top
panels) and CO$_2$ (bottom panels). Experimental results are taken
from Zhou {\it et al} \cite{jila09}.} \label{fig2}
\end{figure}

Next we show in Fig.~2(b) and 2(d) theoretical intensity ratio
$\frac{I_{\perp}}{I_{\parallel}}$ between perpendicular and parallel
components for harmonic orders from H17 to H23, as a function of
alignment angle $\theta$ between pump and probe polarizations.
Experimental results by Zhou {\it et al} \cite{jila09} are shown in
Fig.~2(a) and 2(c) for comparison. For N$_2$, the theoretical
intensity ratio reaches 0.16 at the peak near $55^{\circ}$ for all
the harmonics from H17 to H23 [Fig.~2(b)]. The measurement
\cite{jila09} shows a very similar shape with peaks near
$55^{\circ}$ as well, but the magnitude increasing with harmonic
orders from $0.1$ to $0.22$.  For CO$_2$ [Fig.~2(d)], the shape of
the intensity ratio from theory changes slightly, with the peak now at
about $40^{\circ}$. This is also in good agreement with experiment.
As for the magnitude, the theoretical intensity ratio is about a
factor of three larger than in the experiment. This discrepancy
could be partly due to the fact that near the minimum of the
parallel component [see Fig.~1(b)] the theoretical transition dipole
is not accurate enough or that inner molecular orbitals may
contribute \cite{smirnova09,smirnova09b}. For O$_2$, we found that
the intensity ratio is very small, as expected, with the biggest
intensity ratio of about $1\%$ near $35^{\circ}$.  We comment that
for all three targets, the ratio goes to zero if pump and probe
polarizations are parallel or perpendicular. As stated earlier, this
is expected from symmetry consideration.

\begin{figure}
\centering \mbox{\rotatebox{0}{\myscaleboxc{
\includegraphics{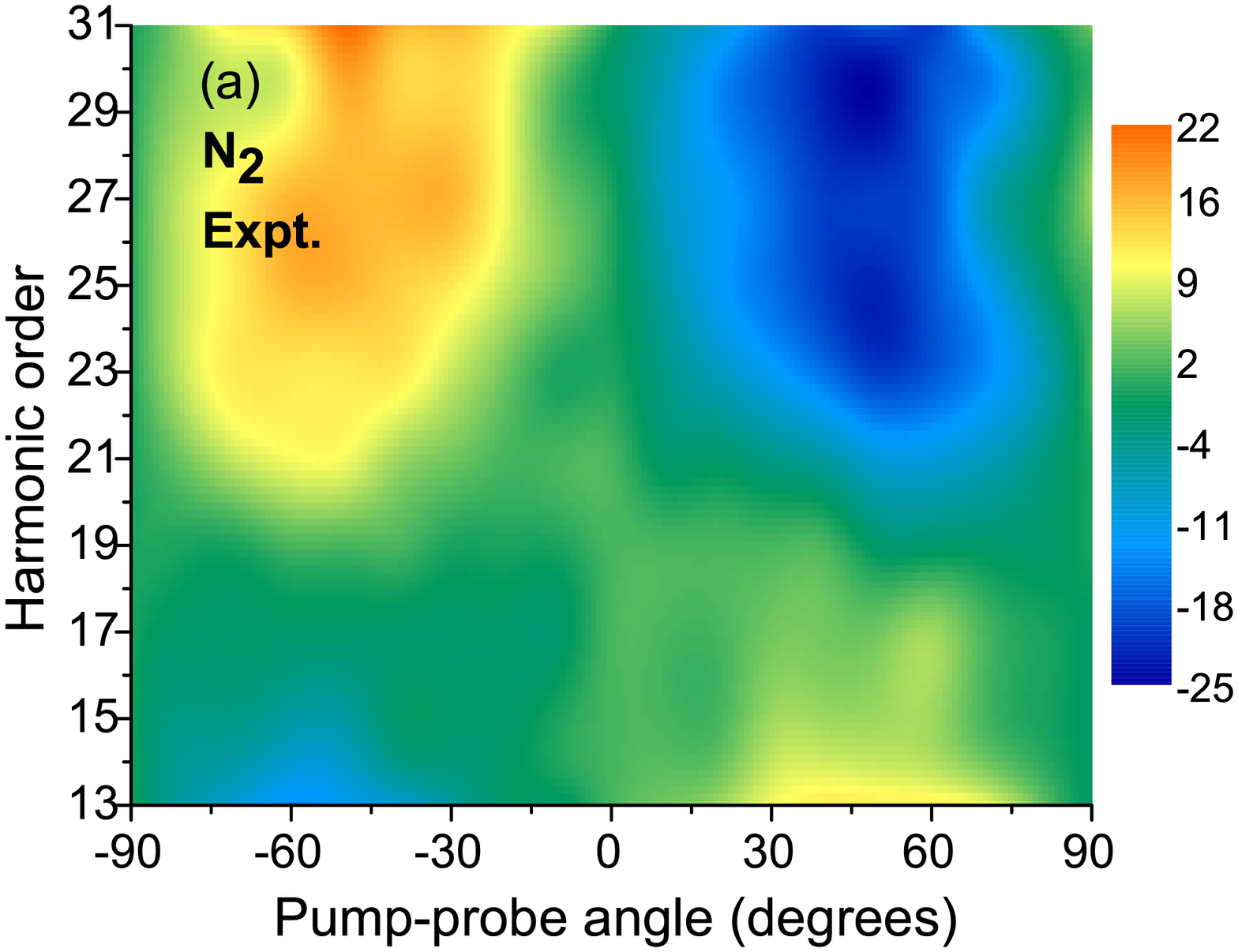}}}}
\centering \mbox{\rotatebox{0}{\myscaleboxc{
\includegraphics{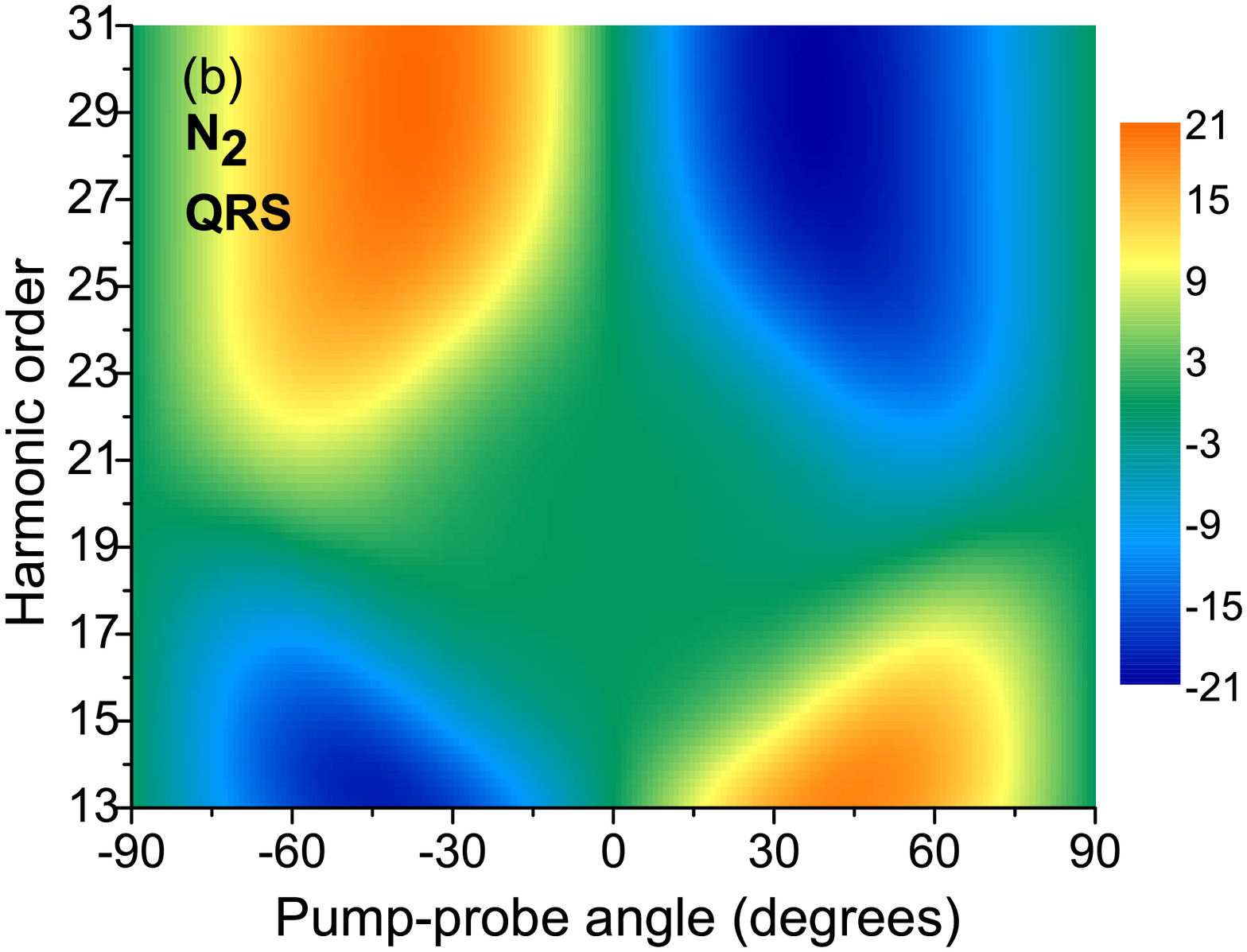}}}}
\centering \mbox{\rotatebox{0}{\myscaleboxc{
\includegraphics{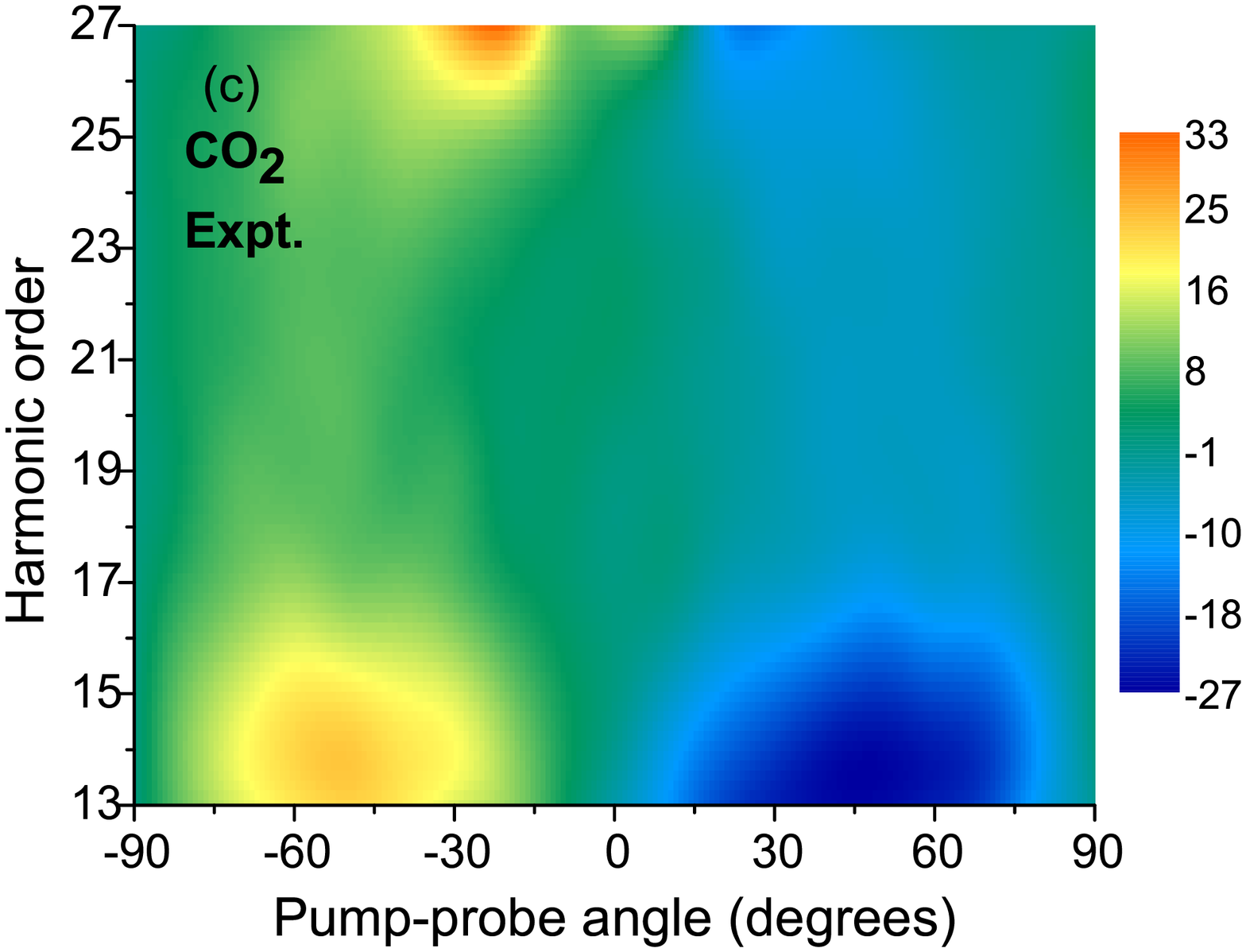}}}}
\centering \mbox{\rotatebox{0}{\myscaleboxc{
\includegraphics{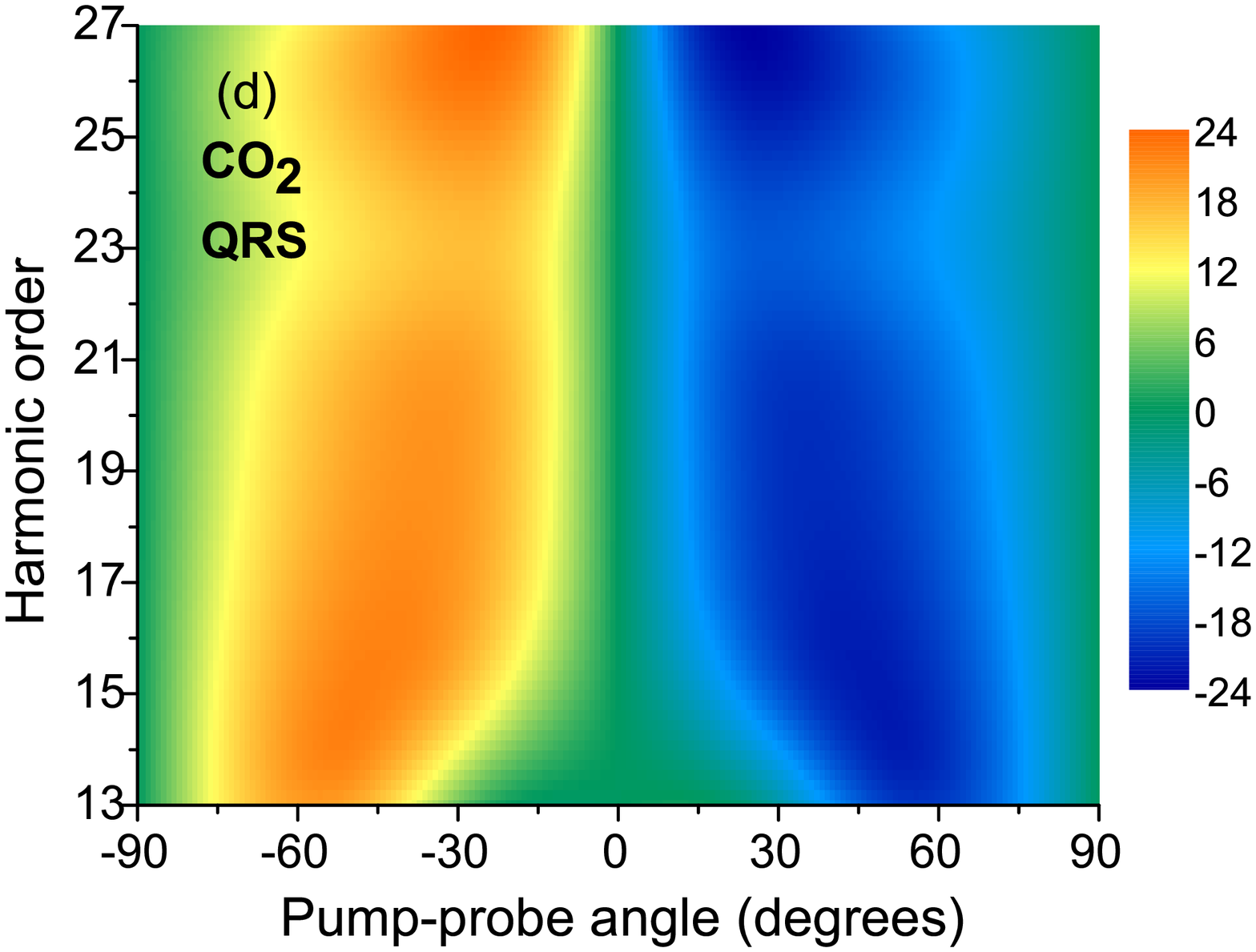}}}}
\centering\mbox{\rotatebox{0}{\myscaleboxc{
\includegraphics{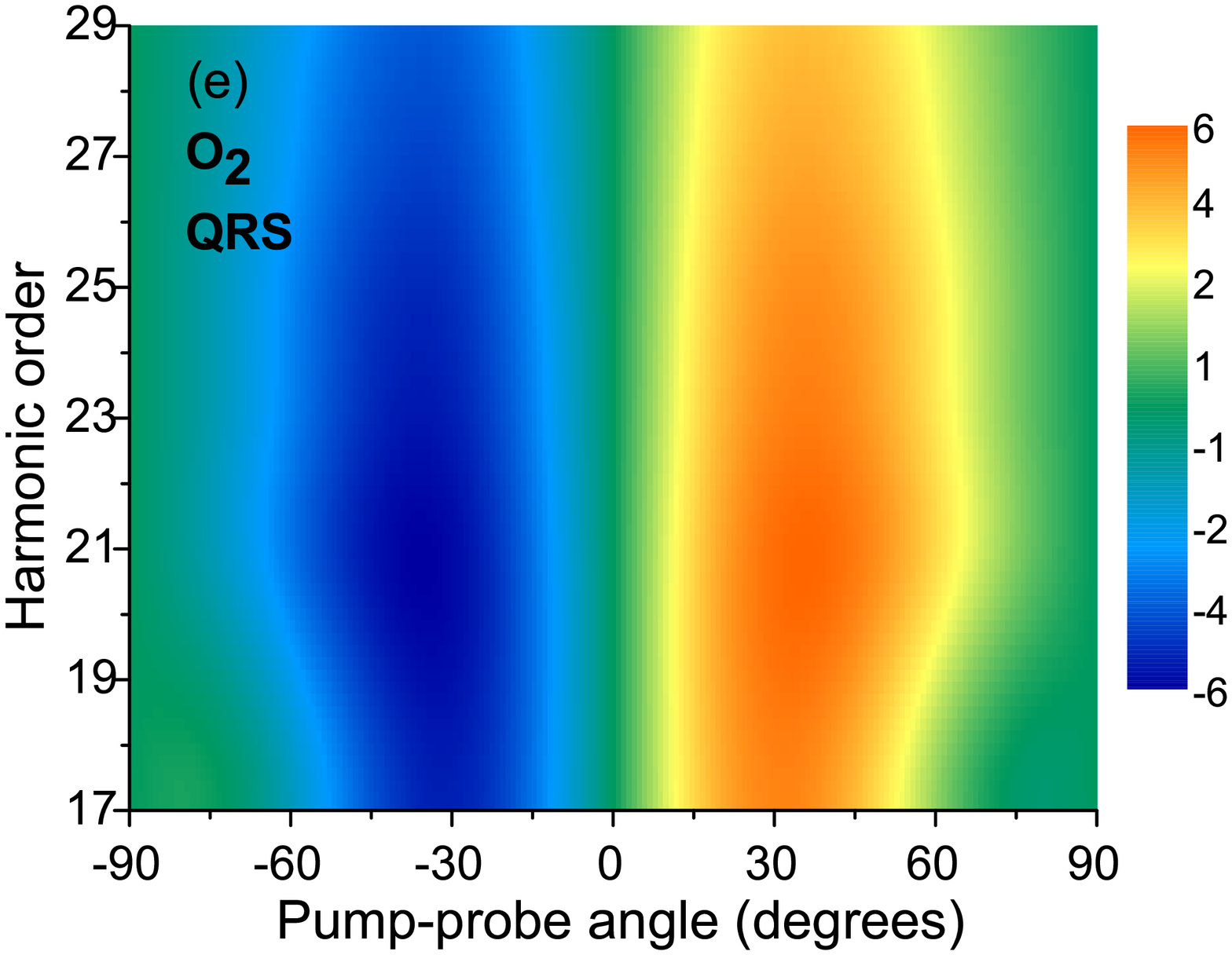}}}}
\caption{(Color online) Experimental (left panels) and theoretical
(right panels) orientation angle $\phi$ (in degrees) as a function
of alignment angle between pump and probe polarization directions
and harmonic order for N$_2$ (top panels), CO$_2$ (middle panels).
Experimental results are taken from Zhou {\it et al} \cite{jila09}.
Theoretical result for O$_2$ is also showed (e).} \label{fig3}
\end{figure}

Let $\delta$ be the phase difference between perpendicular and
parallel components of the harmonic field and $\tan(\gamma)=
\sqrt{\frac{I_{\perp}}{I_{\parallel}}}$. As $\delta \neq 0$ or $\pi$
in general, the emitted harmonic is elliptically polarized. To
characterize the polarization ellipse, we define \cite{born} the
orientation angle $\phi$ of the ellipse and the ellipticity
$\epsilon=\tan(\chi)$ by
\begin{eqnarray}
\tan(2\phi)=\tan(2\gamma)\cos(\delta),\\
\sin(2\chi)=\sin(2\gamma)\sin(\delta).
\end{eqnarray}

Our results for the orientation angle $\phi$ are shown in Fig.~3(b)
and 3(d), as a function of alignment angle between pump and probe
polarizations and harmonic order. Experimental results by Zhou {\it
et al} \cite{jila09} are also shown (left panels) for comparison.
The theoretical data are anti-symmetric with respect to the sign
change in the pump-probe angle $\theta$, so in the following we just
focus on the positive $\theta$. The experimental data are less
symmetric. Here the positive (or negative) angle corresponds to the
case when the major axis of the ellipse rotates in the same (or
opposite) direction as the molecular axis.

The most noticeable feature for N$_2$ is the sign change in the
orientation angle as a function of harmonic order near H19.
The orientation angle of about $20^{\circ}$ at H13, decreases smoothly
with harmonic order, and reaches $-20^{\circ}$ at H29. This is in
excellent agreements with the measurements by Zhou {\it et al},
shown in Fig.~3(a), as well as with Levesque {\it et al}
\cite{levesque07}.  Zhou {\it et al} \cite{jila09} found that the sign of
the orientation angle changes near H19, while Levesque {\it et al}
\cite{levesque07} found the change near H21, independent of the pump-probe
polarization angle. We comment that calculations based on the SFA do
not lead to a satisfactory agreement with experiments
\cite{levesque07}. For CO$_2$, the theoretical orientation angles
are negative (for positive $\theta$) for all the considered
harmonics. This is in good agreements with Zhou {\it et al}, shown
in Fig.~3(c), and Levesque {\it et al} \cite{levesque07}. For O$_2$,
on the other hand, the orientation angle remains positive for the
harmonic range shown in the figure. Its magnitude is also much
smaller, reaching about $6^{\circ}$ near H19-H21 for $\theta\sim
30^{\circ}$ - $40^{\circ}$. This behavior agrees well with Levesque
{\it et al} \cite{levesque07}. Again, these small orientation angles
are due to the small intensity ratios, which in turn is related to
the smallness of the PI cross sections for the perpendicular
component, as compared to the parallel component, for intermediate
angles in O$_2$ [see Fig.~1(c)].

\begin{figure}
\centering \mbox{\rotatebox{0}{\myscaleboxb{
\includegraphics{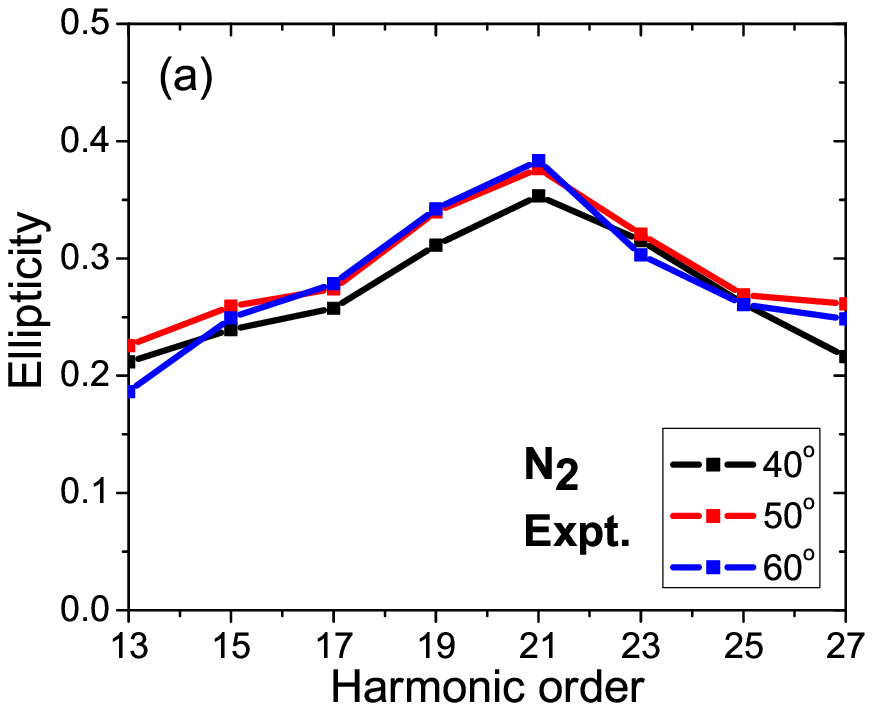}}}}
\centering \mbox{\rotatebox{0}{\myscaleboxb{
\includegraphics{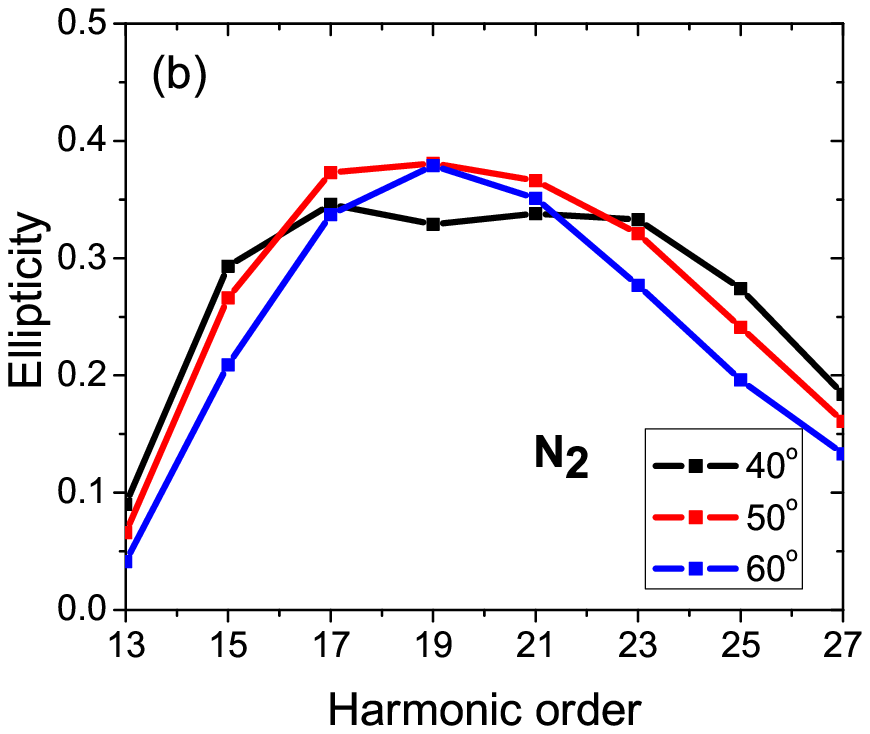}}}}
\centering \mbox{\rotatebox{0}{\myscaleboxb{
\includegraphics{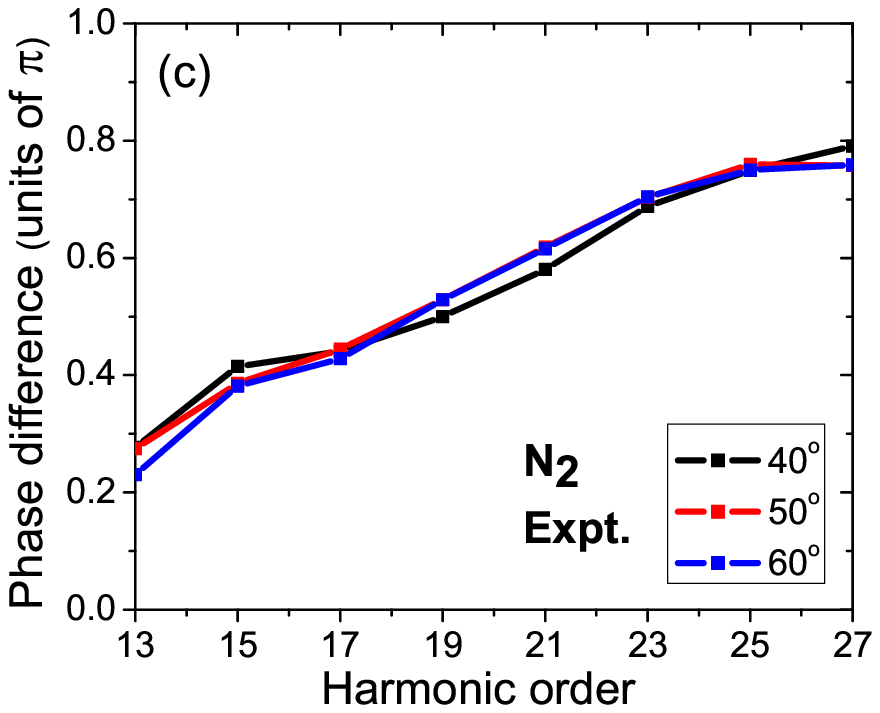}}}}
\centering \mbox{\rotatebox{0}{\myscaleboxb{
\includegraphics{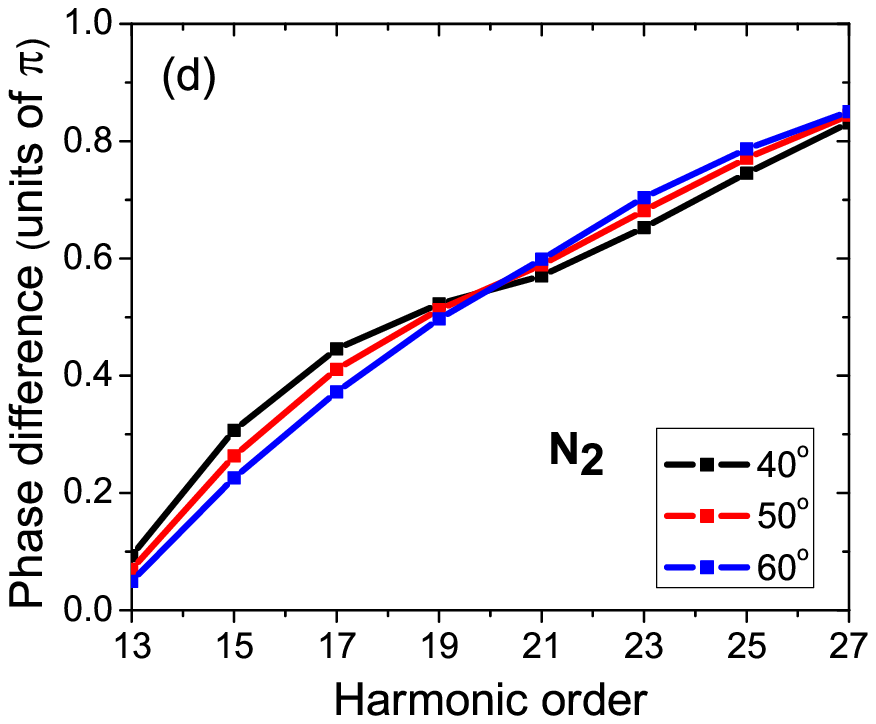}}}}
\caption{(Color online)  Experimental (left panels) and theoretical
(right panels) ellipticity (top panels) and phase difference (bottom
panels) between the two polarization components  from N$_2$ for
angles between pump and probe polarizations $\theta=40^{\circ}$,
$50^{\circ}$, $60^{\circ}$. Experimental results are taken from Zhou
{\it et al} \cite{jila09}.} \label{fig4}
\end{figure}

We also compare ellipticity $\epsilon$ and phase difference $\delta$
vs harmonics order at fixed pump and probe angles
$\theta=40^{\circ}$, $50^{\circ}$, and $60^{\circ}$. Fig.~4 shows
that the theoretical results for N$_2$ are in good agreement with
the experimental data of Zhou {\it et al} \cite{jila09}. In
particular, the theory predicts a large ellipticity up to
$\epsilon\approx 0.4$ near H21, in agreement with experiment. We
comment that a recent calculation based on an extended
stationary-phase SFA by Etches {\it et al} \cite{madsen09} showed
very weak ellipticity of about $0.02$ only. The phase difference, shown
in Fig.~4(d), increases nearly linearly with harmonic order, from
$0.1\pi$ at H13 to $0.8\pi$ at H27, but is nearly independent of
alignment angle. Note that the phase difference is nearly $\pi/2$
at H19. This is exactly the harmonic order, where the orientation
angle changes its sign, see Eq.~(2) and Fig.~3(b). This behavior is
in excellent agreement with experiment shown in Fig.~4(c). For
CO$_2$ the ellipticity from the QRS is slightly smaller than that of
N$_2$, while the measurements by Zhou {\it et al} showed a value of
less than 0.1. This discrepancy is consistent with the larger
errors we found for the CO$_2$ intensity ratio, but the reason
remains largely unclear at this moment. We further note that the
calculation for CO$_2$ by Smirnova {\it el al} \cite{smirnova09}
showed an ellipticity of $0.1$ at H29, which increases up to about
0.4 at harmonics H37-H43. In their simulation, contributions from
two lower molecular orbitals were also included. For completeness we
note that the QRS predicts an ellipticity for O$_2$ of less than
$5\%$ under the same experimental conditions.

In general, experimental HHG spectra include the effect of
macroscopic propagation in the medium \cite{gaarde08}. However,
under typical experimental conditions, we can show that macroscopic
propagation will affect both harmonic components in the same way.
Indeed, the propagation equation for each harmonic component $E_{a}$
(with $a=\parallel$ or $\perp$) can be written under the paraxial
approximation as \cite{gaarde08,jin09}
\begin{eqnarray}
\nabla^{2}_{\perp}E_{a}(r,z,\omega,\theta)-
\frac{2i\omega}{c}\frac{\partial E_{a}(r,z,\omega,\theta)}{\partial
z} \propto \langle D_{a}(r,z,\omega)\rangle_{\theta},
\end{eqnarray}
where $\langle D_{\parallel,\perp}(r,z,\omega)\rangle_{\theta}$ is
the nonlinear polarization, averaged over the molecular alignment
distribution for a fixed pump-probe angle $\theta$. Here we assume
that absorption and free-electron dispersion are negligible. In a
typical gas jet experiment, the aligning laser is much less intense
and more loosely focused than the probe laser. Therefore we can
assume that the aligning laser is uniform in the gas jet, which is
typically of about 1 mm thick. We found that for a fixed
$\{\omega,\theta\}$ the intensity ratio and phase difference between
the two components $\langle D_{\parallel}(\omega)\rangle_{\theta}$
and $\langle D_{\perp}(\omega)\rangle_{\theta}$ change less than
$10\%$ as probe laser intensity changes from $1.5\times 10^{14}$ to
$2.5\times 10^{14}$ W/cm$^2$. In other words, the ratio $R=|\langle
D_{\perp} \rangle /\langle D_{\parallel} \rangle|$ and phase
difference are nearly independent of the spatial coordinates
$\{r,z\}$ in the gas jet. From Eq.~(4), it follows that the ratio
$|E_{\perp}/E_{\parallel}|=R$. The same arguments also hold for the
phase difference between $E_{\perp}$ and $E_{\parallel}$. This
implies that the results presented in this paper should be nearly
unchanged even if the macroscopic propagation is carried out. Our
results are still dependent on the degree of molecular alignment,
which is controlled by the pump pulse. Therefore polarization
resolved HHG measurements allow us to directly extract
single-molecule features (up to averaging over the alignment
distribution) without much influence of the details of the
macroscopic phase-matching conditions.

In conclusion, we have shown that the quantitative rescattering
theory can be extended to calculate polarization and ellipticity of
high-order harmonics from aligned molecules in intense laser fields.
Theoretical results are compared to experimental measurements side
by side and good agreement has been found. The interaction of light
with molecules is governed by the dipole transition matrix elements.
This dipole interaction has been traditionally probed using
photoionization, but can similarly be probed by studying HHG. While
photoionization has the advantage of achieving higher energy
resolution to reveal many-electron dynamics, HHG has the advantage
of surveying a broader photon energy range coherently in one single
experiment, thus revealing the global property of the molecule.
Since the phases of the harmonics can be conveniently measured
experimentally, HHG also has the advantage of revealing directly the
phases of the transition dipoles.

We thank X. Zhou, M. Murnane, and H. Kapteyn for providing us with
their experimental data and the stimulating discussions. This work
was supported in part by the Chemical Sciences, Geosciences and
Biosciences Division, Office of Basic Energy Sciences, Office of
Science, U. S. Department of Energy.  RRL also acknowledges the
support of the Welch Foundation (Houston, TX) under grant A-1020.

\end{document}